\documentstyle[graphicx,natbib,epsfig,amsmath,amssymb,traditabstract,letter]{aa} 

\newcommand{\msun}{\,\hbox{$M_{\odot}$}}

\newcommand{\kms}{\,\hbox{\hbox{km}\,\hbox{s}$^{-1}$}}

\newcommand{\htwo}{\,\hbox{$\rm{H_ 2}$}}

\newcommand{\spi}{{\it Spitzer}}
\newcommand{\ang}{\,\hbox{\AA}}
\newcommand{\neii}{\,\hbox{[\ion{Ne}{II}]}}

\newcommand{\nev}{\,\hbox{[\ion{Ne}{V}]}}
\newcommand{\oi}{\,\hbox{[\ion{O}{I}]}}

\newcommand{\sii}{\,\hbox{[\ion{S}{II}]}}
\newcommand{\oiv}{\,\hbox{[\ion{O}{IV}]}}
\newcommand{\oiii}{\,\hbox{[\ion{O}{III}]}}

\newcommand{\um}{\,\hbox{$\mu$m}}
\newcommand{\vsys}{\,\hbox{$V_{sys}$}}

\begin{document}

   \title{Turbulent and fast motions of H$_2$ gas in active galactic nuclei}

   \subtitle{}

   \author{K. M. Dasyra\inst{1,2} 
          \and
	 F.  Combes\inst{2} 
	  }

   \institute{
	   Laboratoire AIM, CEA/DSM-CNRS-Universit\'e Paris Diderot, Irfu/Service
             dÕAstrophysique, CEA Saclay, F-91191 Gif-sur-Yvette, France\\
         \and
             Observatoire de Paris, LERMA (CNRS:UMR8112), 
             61 Av. de l\'\ Observatoire, F-75014, Paris, France \\
             }

   \date{}
   
    \abstract 
    { Querying the \spi\ archive for optically-selected active galactic nuclei (AGN) observed in high-resolution-mode spectroscopy, we identified radio and/or interacting 
      galaxies with highly turbulent motions of \htwo\ gas at a temperature of a few hundred Kelvin. Unlike all other AGN that have unresolved \htwo\ line profiles at a 
      spectral resolution of $\sim$600, 3C236, 3C293, IRAS09039+0503, MCG-2-58-22 and Mrk463E have intrinsic velocity dispersions exceeding 200 \kms\ for at least 
      two of the rotational S0, S1, S2, and S3 lines.  In a sixth source, 4C12.50, a blue wing was detected in the S1 and S2 line profiles, indicating the presence of a warm 
      molecular gas component moving at -640\kms\ with respect to the bulk of the gas at systemic velocity. Its mass is 5.2$\times$10$^7$ \msun , accounting for more than 
      one fourth of the \htwo\ gas at 374K, but less than 1\% of the cold \htwo\ gas computed from CO observations. Because no diffuse gas component of 4C12.50 has been 
      observed to date to be moving at more than 250 \kms\ from systemic velocity, the \htwo\ line wings are unlikely to be tracing gas in shock regions along the tidal tails 
      of this merging system. They can instead be tracing gas driven by a jet or entrained by a nuclear outflow, which is known to emerge from the west  nucleus of 4C12.50. 
      It is improbable that such an outflow, with an estimated mass loss rate of 130 \msun yr $^{-1}$, entirely quenches the star formation around this nucleus.
      }


   \titlerunning{Turbulent and fast motions of \htwo\ gas in AGN}
   \authorrunning{K. M. Dasyra \& F. Combes}
   
   \maketitle


\section{Introduction}
\label{sec:intro}

Large-scale feedback effects such as jets and outflows from active galactic nuclei (AGN) are thought to be
capable of affecting the formation of new stars in their host galaxies. The triggering of star formation by 
compression of gas \citep[e.g.,][]{vanbreugel85}, as well as the suppression of star formation by heating of gas
that prevents its further collapse \citep[e.g.,][]{nesvadba10} have been observed in local AGN.  Cosmological 
simulations have suggested that AGN feedback effects, which are often associated with mergers, could make 
galaxies appear red, or even explain the observed luminosity functions of  galaxies \citep{croton06,hopkins06}. 
Combined with multi-wavelength observations indicating that the star-formation history and the black-hole-accretion 
history of the Universe peak at comparable redshifts, between 1$\lesssim$$z$$\lesssim$3 \citep{marconi04, merloni04}, 
this suggests that AGN feedback could have affected the shape of present-day galaxies considerably.

Extensive tests of the role of AGN feedback on the interstellar medium (ISM) of galaxies require 
a detailed kinematic study of outflowing gas in local sources. Signs of massive gas outflows have been 
detected for ionized atomic gas \citep[e.g.,][]{veilleux95,emonts05,holt06,muellersanchez06}, neutral
atomic gas  \citep[e.g.,][]{morganti05,rupke05}, and molecular CO and OH gas \citep[e.g.,][]{curran99,
das05,garciaburillo09,sakamoto09,feruglio10,fischer10,sturm11}. In this letter we present evidence 
for the first detection of highly turbulent motions of \htwo\ gas at a temperature of a few hundred Kelvin
as seen with \spi\ for several local AGN. We adopt H$_0$=70 \kms\  Mpc$^{-1}$, $\Omega_{M}$=0.3, 
and $\Omega_{\Lambda}$=0.7 throughout this work.


\begin{figure*}
\begin{center}
\includegraphics[width=18cm]{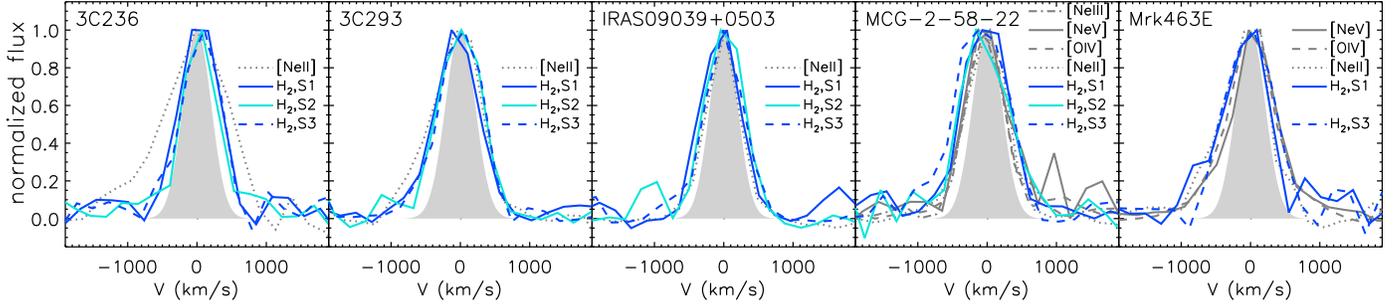} 
\caption{Normalized, continuum-subtracted molecular and ionized gas line profiles in local AGN with resolved \htwo\ emission.
The filled gray area represents a Gaussian with FHWM equal to the resolution of the IRS.}
\label{fig:profiles}
\end{center}
\end{figure*}

\begin{table*}
\caption{\label{table:line_properties} Fluxes and widths of resolved \htwo\ lines in local AGN.}
\centering
\scalebox{0.84}{
\begin{tabular}{ p{2.4cm} p{0.7cm} c c c c c c c c c }
\hline\hline
Source & $z$ & f$_{S1}$ & FWHM$_{S1}$\tablefootmark{a}  & f$_{S2}$ & FWHM$_{S2}$\tablefootmark{a}  & f$_{S3}$ & FWHM$_{S3}$\tablefootmark{a} & M$_{warm}$ & T & M$_{cold}$\tablefootmark{b}\\
- & - & 10$^{-17}$W\ m$^{-2}$ & \kms  & 10$^{-17}$W\ m$^{-2}$ & \kms  & 10$^{-17}$ W\ m$^{-2}$ & \kms  & 10$^7$\msun  &  K & 10$^9$\msun  \\
\hline
3C236                 		& 0.0989  & 1.06$\pm$0.07 & 582 (750) & 0.49$\pm$0.08 & 557 (719) & 0.81$\pm$0.09 & 624 (767) & 6.10 & 345 & $<$5.1 \\
3C293                     		& 0.0450  & 5.30$\pm$0.38 & 510 (716) & 1.79$\pm$0.31 & 553 (732) & 2.96$\pm$0.38 & 519 (714) & 6.62 & 323 & 23 \\
IRAS09039+0503 		& 0.1254   & 3.09$\pm$0.54 & 519 (700) & 1.34$\pm$0.06 & 575 (730) & 2.04$\pm$0.13 & 569 (751) & 31.6 & 334 & ... \\
MCG-2-58-22        		& 0.0472 & 3.86$\pm$0.42 & 544 (740) & 1.50$\pm$0.25 & 636 (796) & 3.40$\pm$0.46 & 662 (824) & 4.19 & 359 & 5.7 \\
Mrk463E                 		& 0.0507 & 3.37$\pm$0.59 & 556 (748) & $<$1.98 & ... (...) & 3.76$\pm$0.44 & 590 (767) & 4.06 & 378 & 1.2 \\
4C12.50 (main\tablefootmark{c}) & 0.1217 & 1.64$\pm$0.21 & ... (519) & 1.20$\pm$0.11 & ... (552) & 1.78$\pm$0.46 & ... (652) & 13.9 & 374 & 15 \\
4C12.50 (wing\tablefootmark{c}) & 0.1196\tablefootmark{d} & 0.62$\pm$0.21 & ... (568)  & 0.44$\pm$0.11 & 521 (690) & ... & ... (...) & 5.19 & 374\tablefootmark{e} & ...\\
\hline
\end{tabular}}
\tablefoot{None of these sources had a reliable S0 detection owing to the high noise levels in the long wavelength array.
\tablefoottext{a}{Rest-frame, instrumental-broadening-corrected FWHM.  Their measured values appear in parenthesis. Their error bars 
correspond to 5$-$10\% of their values.}
\tablefootmark{b}{  Cold \htwo\ gas masses calculated from CO observations \citep{evans99, evans02,bertram07,saripalli07} 
using an \htwo / CO mass conversion factor of 1.5\msun /(K \kms pc$^2$) for the IR-bright systems Mrk463E and 4C12.50  \citep{evans02},
and a standard Galactic value of 4.8\msun /(K \kms pc$^2$) for all other sources. They are converted to the adopted cosmology. }
\tablefoottext{c}{ The main and wing components correspond to the primary and the secondary Gaussian functions in Figure~\ref{fig:4c12.50spec}. }
\tablefoottext{d}{ The Gaussians that best fit the S1 and S2 line wings peak at -646 \kms\ and -634 \kms\ from systemic velocity, respectively. }
\tablefoottext{e}{ The excitation temperature of the rapidly moving \htwo\ gas is assumed to be identical to that of the \htwo\ at systemic velocity to facilitate
the comparison of their masses. }
}
\end{table*}
  
\section{The sample selection}
\label{sec:data}

We queried for turbulence in the warm \htwo\ gas in local AGN using mid-IR spectra obtained with 
\spi\ in high-resolution mode. The full archival sample comprises 298 sources that are classified as AGN based 
on optical spectroscopy. It is presented in \citet{dasyra11} together with the data reduction techniques and the
extracted spectra.

To look for turbulent \htwo\ gas motions, we examined the profiles of the purely rotational (0-0)S0 28.22\um , (0-0)S1 
17.04\um , (0-0)S2 12.28\um , and (0-0)S3 9.66\um\  lines. We searched for either resolved profiles with velocity dispersion 
$\sigma$$\gtrsim$200 \kms, or for profiles with asymmetric wings that are characteristic of outflows. To ensure the 
reliability of our results, we only examined sources with at least two lines of signal-to-noise (S/N) ratio $>$5.  We also 
requested that at least two lines suggest a similar kinematic pattern, i.e. a wing or a resolved profile. To consider a line 
resolved we requested that its full width at half maximum (FWHM) value minus the FWHM  error exceeds the instrumental 
resolution $R$ plus the resolution error at the observed-frame wavelength of the line \citep{dasyra11}. The average $R$ 
value in the 12.0$-$18.0\um\ range, which comprises the \htwo\ S1 and S2 transitions, is 507$\pm$66 \kms.

\section{Results: Sources with highly turbulent or rapidly moving \htwo\ gas in the warm phase}
\label{sec:results}

Of the 298 sources 62 had at least two \htwo\ lines detected with S/N$>$5. The profiles of two or more lines were spectrally 
resolved in only five sources, namely 3C236, 3C293, IRAS09039+0503, MCG-2-58-22, and Mrk463E (Fig.~\ref{fig:profiles}). 
Their velocity dispersions are in the range 220$\lesssim$$\sigma$$\lesssim$280\kms\ (Table~\ref{table:line_properties}). 
Because all these sources are radio galaxies and/or interacting systems, the high turbulence in their warm \htwo\ motions can 
be driven by AGN feedback mechanisms, by gravitational instabilities, or by supernova winds. Still, no mechanism is efficient 
enough to kinematically distort a detectable mass of warm \htwo\ gas to velocity dispersions exceeding 300\kms .  We computed 
the excitation temperature T and mass of the turbulent gas (Table~\ref{table:line_properties}) using the detected S1, S2, and S3 
line fluxes as in \citet{higdon06}. At temperatures of 300$-$400 K, its mass is typically on the order of 1\% of the cold \htwo\ gas 
mass indicated by CO observations.

Further outflow or inflow signatures were sought for in the \htwo\ line wings and in the difference of the \htwo\ recession velocity from the 
systemic velocity, \vsys . The latter was determined from the \neii\ line, emitted by ions that are abundant in star-forming regions and in 
the AGN vicinity owing to their low ionization potential, 21.56 eV. The \htwo\ recession velocity agreed within the errors with \vsys\ for all 
sources, including those with massive outflows of the gas that is photoionized by the AGN and that is traced by the \nev\ or \oiv\ lines 
\citep[i.e., 3C273, IRAS13342+3932, IRAS05189-2524, IRAS15001+1433, IRAS23060+0505, Mrk609;][]{dasyra11}.


\begin{figure}[h!]
\begin{center}
\includegraphics[width=8.4cm]{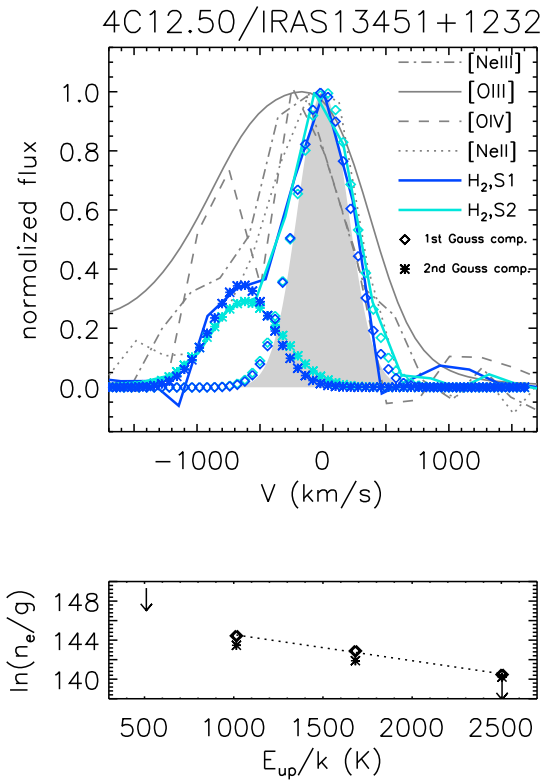} 
\caption{Upper panel: Normalized, continuum-subtracted molecular and ionized gas line profiles for 4C12.50.  In addition to the profiles of
the MIR lines, the 5007\ang\ \oiii\ profile that is convolved to the resolution of IRS is presented for comparison. The blue wing is detected in all 
lines except for the unresolved S3. Lower panel: \htwo\ excitation diagram constructed separately for the primary Gaussian component 
(open diamonds) and the secondary Gaussian component (stars) of each \htwo\ line.
}
\label{fig:4c12.50spec}
\end{center}
\end{figure}


\begin{figure}[h!]
\begin{center}
\includegraphics[width=6.3cm]{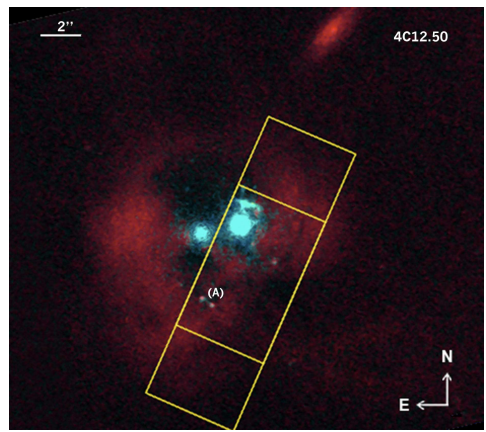} 
\caption{ Composite H$\alpha$ (cyan) and 5900\ang\ continuum (red) image of 4C12.50, constructed from {\it Hubble} Space Telescope data \citep{batcheldor07}. 
To enhance the visibility of low surface brightness structures, we removed all symmetric galaxy components (i.e., a common bulge and two residual disks located at 
the position of the two nuclei) from the 5900\ang\ image with GALFIT \citep{peng02}. Both disks were found to have an inclination of $\sim$20$^{\circ}$. Irregularly 
moving components are found within the IRS slit, marked with a yellow box for both nod positions. (A) marks the position of two super star clusters with velocities that 
are blueshifted by up to $\sim$250 \kms\ from systemic velocity \citep{zaurin07}. 
}
\label{fig:4c12.50slits}
\end{center}
\end{figure}


The only source with wing signatures in its \htwo\ line profiles was 4C12.50, also known as IRAS13451+1232 or PKS1341+12. The S1 and S2 line wings 
(Fig.~\ref{fig:4c12.50spec}; upper panel), detected with S/N$\gtrsim$3, point at two warm molecular gas kinematic components 
in this source. The peak of the secondary Gaussian that is needed to fit both profiles is blueshifted by $\sim$640\kms\ from the primary Gaussian at \vsys .
The flux ascribed to the primary Gaussian is only 2.6 and 2.7 times higher than that ascribed to the secondary Gaussian of the S1 and S2 transitions, respectively. 
Using the actual flux in each Gaussian (Table~\ref{table:line_properties}), we separately calculated the \htwo\ gas mass for each kinematic component. We 
find the excitation temperature of the bulk of the gas to be 374$\pm$12 K. It is equal to the inverse of the slope of the line that best fits the excitation 
diagram points \citep{rigopoulou02}. The excitation diagram (Fig.~\ref{fig:4c12.50spec}; lower panel) shows the natural logarithm of the number of electrons  
n$_e$ that descended from the upper to the lower state, normalized by the statistical weight g of the transition, as a function of the temperature that corresponds 
to the energy of the upper state E$_{up}$ divided by the Boltzmann constant $k$. The value of n$_e$ is computed as \hbox{L/($\alpha  h \nu$)}, where h is the Planck 
constant, $\alpha$ is the Einstein coefficient of the transition, and $\nu$ is the frequency of the emitted line. For a single temperature of 374K we find that the mass 
of the \htwo\ gas at systemic velocity is 1.4$\times$10$^8$\ \msun\ \citep[see also][]{higdon06}. Assuming (for simplicity) that T is the same for both \htwo\ kinematic
components (see Figure~\ref{fig:4c12.50spec}), we find that the mass of the \htwo\ gas moving at 640\kms\ is  5.2$\times$10$^7$\ \msun . This is  0.3\% of the 
cold \htwo\ gas mass in the west nucleus of 4C12.50, which was found from CO observations to be 1.5$\times$10$^{10}$\msun\ \citep{evans02}. 

\section{Discussion: An AGN-driven molecular gas outflow in 4C12.50?}
\label{sec:discussion}

Even though 4C12.50 is an IR-bright system of two interacting galaxies \citep{axon00}, the secondary Gaussian of Figure~\ref{fig:4c12.50spec} is unlikely to be tracing 
gas in the east nucleus, which is mostly located outside the IRS slit (Fig.~\ref{fig:4c12.50slits}). Any residual gas from the east nucleus inside the slit would be moving 
at a velocity comparable to the difference in the recession velocity of the two nuclei, $\sim$200\kms . This difference is computed from ionized gas kinematics \citep{holt03}, 
and it is confirmed by stellar kinematics from CO absorption features presented in \citet{dasyra06}. Besides the two nuclei residing in a common bulge, 4C12.50 also 
has off-nuclear gas concentrations and super star clusters in tidal tails. The blueshifted \htwo\ emission could arise from gas in a tidal tail inside the IRS slit 
(Fig.~\ref{fig:4c12.50slits}), which is entering a shock front created during the collision \citep[see][]{cluver10}. The tidal tail could be moving faster or be at a different inclination
angle $i$ than its corresponding nucleus, of $\sim$20$^{\circ}$ in either case. Because the deprojected tail velocity would be equal to 640/$sin(i)$ \kms , it could reach 
a value as high as $\sim$2000\kms  . This scenario is also unlikely given that no off-nuclear, large-scale ($<$20 kpc) kinematic component has been observed to be moving 
faster than $\pm$250\kms\ from \vsys\ along the line of sight \citep{holt03,zaurin07}. 

A scenario that agrees better with existing observations is that the \htwo\ gas is moving toward us driven by feedback mechanisms \citep[e.g.,][]{alatalo11}.
Optical spectroscopy indicated the presence of an outflow from the west nucleus of 4C12.50 by revealing the existence of three kinematic components for the
nuclear \oi , \sii , and \oiii\  emission \citep{holt03}. Most of the \oiii\ emission is blueshifted by 400\kms , while its broadest component (of FWHM$\sim$1900 \kms ) 
is blueshifted by 2000\kms\ from \vsys . MIR spectroscopy suggested an AGN-related nuclear outflow of ionized gas. Blue wings were observed in the profiles of
the \oiv\ 25.89\um\ and the \nev\ 14.32\um\ lines \citep{spoon09,dasyra11}, emitted by ions that are primarily found in hard radiation fields. Radio observations 
revealed an HI absorption line blueshifted by $\sim$1000 \kms\ from \vsys\ \citep{morganti04}. Because a background radio source is required for HI absorption 
to be seen, the hydrogen clouds are likely to be located between the observer and the AGN or its jet. Estimates of the outflow mass range from 8$\times$10$^5$\msun\  
for the ionized 10$^4$K gas \citep{holt11} to 5.6$\times$10$^8$\msun\  for the neutral gas traced by \ion{Na}{I}D \citep{rupke05}, bracketing our mass 
estimate, $M_{out}$, of 5.2$\times$10$^7$\msun\ for the outflowing $\sim$400K \htwo\ gas.

If the outflow is caused by AGN radiation pressure or winds \citep{holt11}, it can be considered spherical. If the gas is also distributed in a sphere, and if its density 
is falling with the inverse square of the distance from the center, its mass outflow rate, $\dot{M}$, will be given from the product \hbox{$M_{out} V_{out} R^{-1}$}.
For an outflow velocity $V_{out}$ of 640\kms\ and for a radius $R$ of 270 pc, as estimated for the CO gas assuming that it is thermalized to the dust temperature 
\citep{evans02} and as converted to the adopted cosmological distance,  $\dot{M}$ will be 130 \msun yr$^{-1}$.  An outflow of these properties is unlikely to entirely 
suppress star formation in the west nucleus of 4C12.50, whose star-formation rate (SFR) is estimated to be between 370-1380 \msun yr$^{-1}$. The lower value 
is found from the CO mass using a gas consumption timescale of 4$\times$10$^7$yrs, while the upper value is found by folding the CO-based \htwo\ mass and 
radial extent in the Schmidt-Kennicutt law \citep{evans02}.  If the outflow were symmetric, both a blue and a red \htwo\ 
line profile wing should exist unless the gas moving away from the observer is obscured by dust. This is plausible for a source with E(B-V) of 1.44 magnitudes 
\citep{holt11} and 9.7\um\ optical depth of 0.59 \citep{veilleux09}, which could translate into a 10$-$20\% absorption of the total flux at 17\um\ \citep{li01}, and 
which could preferentially suppress the red line wing for a circumnuclear dust distribution.

Alternatively, a radio jet encountering clouds on its path could be driving the outflow \citep[e.g.,][]{dietrich98}. A jet is indeed known to emerge from the west nucleus 
of 4C12.50. It extends out to 45 pc in projection in the north and 170 pc in the south \citep{stanghellini97}, and it propagates close to the speed of light at a small 
angle from the line of sight  \citep{lister03}.  A previous flare of the jet, undetected in the radio because of its weak signal, could be associated with the shocked gas 
whose extended X-ray emission peaks at 20 kpc south of the nucleus \citep{siemiginowska08}. The scenario of a jet-driven outflow can easily explain the observed 
line profiles. The detection of a blue or a red wing is random since it depends on the location of the clouds with respect to the jet propagation axis.

\section{Summary and concluding remarks}
\label{sec:conclusions}
We queried the archival catalog of 298 optically-selected AGN observed with \spi\ IRS in high-resolution mode \citep{dasyra11}, aiming to identify sources with 
turbulent motions of their warm molecular gas. We examined the profiles of the  \htwo\ S0, S1, S2, and S3 lines and found only five radio and/or interacting 
galaxies with $\sigma$$>$200 \kms\ but no source with $\sigma$$>$300 \kms . In a sixth source, 4C12.50, the S1 and S2 lines have a blue wing that points 
at warm gas moving toward us with 640\kms . Its mass, 5.2$\times$10$^7$\ \msun , corresponds to an impressively high fraction, $\sim$1/4, of the total 
$\sim$400K \htwo\ gas mass. While it could be tracing shock regions along tidal tails, it is more likely to be tracing an AGN jet or wind-driven outflow, known 
from ionized and neutral gas kinematic studies. Even if all of this gas is entrained by an outflow, it is unlikely to entirely suppress star formation in 4C12.50.  
Additional tests of the role of AGN feedback mechanisms in increasing the turbulence of the molecular gas require observations of high-rotational-number transitions 
of CO molecules that can be mostly excited by the AGN \citep{vanderwerf10}. An essential role in revisiting this question will also be played by the Atacama Large 
Millimeter Array, which, via high-resolution studies of the cold gas, will enable a comparison never performed before: the computation of the warm-to-cold molecular 
gas mass ratio in an outflow vs the rest of the ISM.


\begin{acknowledgements}
 K. D. acknowledges support by the European Community through a Marie Curie Fellowship (PIEF-GA-2009-235038) 
 awarded under the 7th Framework Programme (2007-2013).
\end{acknowledgements}


{}

\end{document}